\documentclass[aps,prl,twocolumn,showpacs,superscriptaddress,groupedaddress]{revtex4}
\usepackage{mathrsfs}  
\usepackage{graphicx,amsmath,bm,subfigure,amssymb,CJK,float}

\begin{document}
\title{Controlling Fusion of Majorana Fermions in one-dimensional systems by Zeeman Field}
\author{L. B. Shao}
\email{lbshao@nju.edu.cn} \affiliation{ National Laboratory of
Solid State Microstructures and Department of Physics, Nanjing
University, Nanjing 210093, China} \affiliation{ Department of
Physics and Center of Theoretical and  Computational Physics,
University of Hong Kong, Pokfulam Road, Hong Kong, China}
\author{Z. D. Wang}
\email{zwang@hku.hk}
\affiliation{ Department of Physics and Center of Theoretical  and
Computational Physics, University of Hong Kong, Pokfulam Road, Hong
Kong, China}
\author{R. Shen}
\affiliation{ National Laboratory of Solid State Microstructures and
Department of Physics, Nanjing University, Nanjing 210093,
China}\author{L. Sheng} \affiliation{ National Laboratory of Solid
State Microstructures and Department of Physics, Nanjing University,
Nanjing 210093, China}
\author{B. G. Wang}
\affiliation{ National Laboratory of Solid State Microstructures and
Department of Physics, Nanjing University, Nanjing 210093, China}
\author{D. Y. Xing}
\affiliation{ National Laboratory of Solid State Microstructures and
Department of Physics, Nanjing University, Nanjing 210093, China}

\begin{abstract}
We propose to realize Majorana fermions (MFs) on an edge of a
two-dimensional topological insulator in the proximity with $s$-wave
superconductors and in the presence of transverse exchange field
$h$.  It is shown that there appear a pair of MFs localized at two
junctions and that a reverse in direction of $h$ can lead to
permutation of two MFs. With decreasing $h$, the MF states can either be fused
 or form one Dirac fermion on the
$\pi$-junctions, exhibiting a topological phase transition. This
characteristic can be used to detect physical states of MFs when they are transformed
into Dirac fermions localized on the $\pi$-junction. A condition
of decoupling two MFs is also given.
\end{abstract}
\maketitle

In recent several years, how to realize, manipulate, and detect
Majorana fermions (MFs) is one of the most active topics of research
in condensed matter physics.~\cite{fu1,nagaosa,been1,palee1,sau}.
The nonabelian character of the MF makes it to be a promising candidate
for topological quantum computation~\cite{ivanov,kitaev,stern}.
There are many systems that manifest the MF, such as half-quantum
vortices of $p$-wave superconductors~\cite{ivanov,volovik1,volovik},
the hexagonal spin lattice model~\cite{kitaev}, the one-dimensional
(1D) $p$-wave lattice~\cite{kitaev1}, the topological surface state
with proximity to an $s$-wave superconductor~\cite{fu1,qi1},
ultracold atom systems~\cite{sato,zhu} and so on. Since Majorana
bound states are superpositions of electrons and holes in the middle
of superconducting gap, they are neutral zero-energy excitations,
and the particle-hole symmetry causes the antiparticle of an MF to
be itself in the field-theory framework~\cite{jackiw}. Many
proposals have been suggested to explore novel properties of MFs,
such as the electrically detected Majorana
interferometry~\cite{been1}, the Andreev reflection induced by
MFs~\cite{palee1}, the charge transport with Majorana edge
modes~\cite{fu2}, and the teleportation by Majorana bound
state~\cite{fu3}. For a superconducting system only Cooper pairs can
be created and annihilated, and so nonabelian statistics of the MFs
can only be formulated in subspaces of same fermion
parity~\cite{ivanov}. Also, manipulating MFs in 1D systems can be
achieved by using assistant quantum systems such as Coulomb
blockaded quantum dots~\cite{flens} and semiconducting wire networks
composed of trijunctions~\cite{fisher}. Although it has been
reported recently that signature of experiment supports the
existence of MFs~\cite{zuo}, how to detect MFs still remains as an
open question.

In this paper, we propose to realize the MFs on the edge of a 2D
topological insulator. Superconducting order parameters with
amplitude $\Delta_0$ are introduced to the edge states by proximity
effect of $s$-wave superconducting junctions, and a local transverse
Zeeman field $h$ is also applied there. It is found that existence
of a pair of MFs depends explicitly upon the relative magnitude of
$h$ and $\Delta_0$~\cite{black}. As $\mid h\mid>\Delta_0$, two MFs in different
spin components emerge at the two junctions, respectively, and an
inverse Zeeman field will lead to permutation of MFs.   At $\mid
h\mid=\Delta_0$, there will be a topological transition. For $\mid
h\mid<\Delta_0$, the MFs can either be fused or form Dirac fermions localized on the junctions, depending on the
phase differences of the junctions. When the phase difference is
unequal to $(2N+1)\pi$, the wavefunctions of MFs are extended into the bulk and fused;
otherwise, one additional MF is created at the junction and combined
with the original one to form one Dirac fermion. Therefore, when the
phase difference of only one junction is equal to $(2N+1)\pi$, the
MF on the junction is effectively driven to the other junction and
forms one Dirac fermion there. This character can be used to detect physical states of
two MFs. The coupling between two MFs will vanish under some
conditions. In the present proposal, all the processes allow us to
realize, manipulate, and detect the MFs readily.
\begin{figure}[hbtp!]
\includegraphics[width=2in]{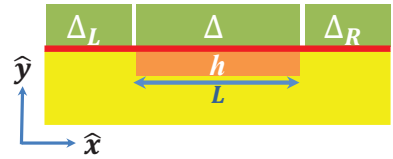}
\caption{(Color online) A half-infinite 2D topological insulator
(yellow region) and its 1D edge (red line) along the $\hat{x}$
direction. Three blocks of superconductors (green blocks) are
fabricated on its top to form two superconducting junctions at $x=0$
and $x=L$. A local Zeeman field along the $\hat{y}$ direction is
applied to the middle (brown) region of $0<x<L$. } \label{model}
\end{figure}

It has been shown theoretically and experimentally that the quantum
spin Hall effect can be realized in HgTe/(Hg,Cd)Te quantum wells and
the gapless edge state is protected by time reversal
symmetry~\cite{bernvg,konig}. In Fig.~\ref{model}, we consider a
half-infinite 2D topological insulator, and its edge is in the
proximity to three blocks of  $s$-wave superconductors with
different phases.  The Hamiltonian of this system may be written as
\cite{been1}
\begin{eqnarray}\label{h}
H&=&\int dx\{\psi_{\alpha}^{\dagger}(x)
(\hat{p}\sigma_{1}+h(x)\sigma_{2}-\mu)_{\alpha\beta}
\psi_{\beta}(x)\nonumber\\&\quad&+\Delta\psi^{\dagger}_{\uparrow}(x)
\psi^{\dagger}_{\downarrow}(x)
+\Delta^{*}\psi_{\downarrow}(x)\psi_{\uparrow}(x)\}. \label{h}
\end{eqnarray}
Here the first term is the Hamiltonian of the 1D edge state with a
uniform Zeeman field $h$ along the the $y$ direction for $0<x<L$, in
which $\hat{p}=-i\partial_{x}$ is the momentum operator,  $\sigma_i$
are the spin Pauli matrices, $\mu$ is the Fermi energy. Also, $\hbar=1$s
and Fermi velocity $\upsilon_{F}\equiv1$ have been taken.
Hamiltonian (\ref{h}) is not invariant under time reversal because
of the presence of Zeeman field $h$. The superconducting order
parameter is given by
\begin{equation}\label{scop}
\Delta(x)=\Bigg\{\begin{array}{c}
\Delta_{0}e^{i\varphi_{L}}\qquad\quad x<0, \\ \;\:
\Delta_{0}\qquad\qquad
0<x<L,\\\Delta_{0}e^{i\varphi_{R}}\qquad\quad x>L ,
\end{array}
\end{equation}
as shown in Fig.~\ref{model}, with $\Delta_{0}>0$ and $\varphi$'s as
the phase of each superconducting region. Obviously, the charge
conjugation symmetry is preserved. The quasiparticle operator in the
Nambu representation
$|\Psi(x)\rangle=[\psi_{\uparrow}(x),\psi_{\downarrow}(x),
\psi_{\downarrow}^{\dagger}(x),-\psi_{\uparrow}^{\dagger}(x)]^{T}$
is defined as $\gamma=\int
dx\{u_{\uparrow}^{*}\psi_{\uparrow}(x)+u_{\downarrow}^{*}\psi_{\downarrow}(x)
+v_{\downarrow}^{*}\psi_{\downarrow}^{\dagger}(x)-v_{\uparrow}^{*}\psi_{\uparrow}^{\dagger}(x)\}.$
When the quasiparticle has relation $\gamma=\gamma^{\dagger}$, it is
a neutral MF. The fact that quasiparticle annihilates itself leads
to $u_{\uparrow}=-v_{\uparrow}^{*}$ and
$u_{\downarrow}=v_{\downarrow}^{*}$. By calculating the equation of
motion given by $E\gamma=[\gamma,H]$, we recover the form of
Bogoliubov-de Gennes (BdG) Hamiltonian in $(u_{\uparrow}, u_{\downarrow}, v_{\downarrow}, v_{\uparrow})^{T}$ as
\begin{eqnarray}
\mathcal{H}_{BdG}=\hat{p}\sigma_{1}\tau_{3}+h(x)\sigma_{2}-\mu\tau_{3}
+\mathbf{Re}\Delta\tau_{1}-\mathbf{Im}\Delta\tau_{2}. \label{bdg}
\end{eqnarray}
Here $\tau$ is the Pauli matrices in the Nambu representation. It
has been pointed out that the Dirac field with the $s$-wave
superconducting order parameter is equivalent to that of a $p+ip$
superconductor that has zero-mode MFs \cite{lutchyn,oreg,qi_rmp}. In Eq.
(\ref{bdg}), the existence of zero modes  relies only on vanishing
of the determinant for Hamiltonian (\ref{bdg}). As a result, the
wave vector is readily solved as $k=\pm\sqrt{\mu^{2}-h^{2}}\pm
i\Delta_{0}$ in the middle region. The imaginary wave vectors
indicate that there are localized states that may give rise to the
MFs. It  should be noted that the first term of Hamiltonian
(\ref{h}) in the absence of $\Delta_0$ breaks the time reverse symmetry
 and yields two Fermi zero modes localized at $x=0$ and $x=L$,
which decay as $e^{-|h|x}$. If we choose that superconducting order
parameters are introduced at $\mu=0$, only two Fermi zero modes have
contribution to superconducting condensation. When $\mu=0$,
the continuous spectrum of Eq. (\ref{bdg}) in the middle region is given
by $E=\sqrt{k^{2}+(h\pm\Delta_{0})^{2}}$ with two energy gaps
$\Delta_{\pm}=\mid h\pm\Delta_{0}\mid$. Obviously, the topological
phase transition happens when the gap is closed at
$h=\pm\Delta_{0}$.

We first focus on the left junction at $x=0$ in the case of
$h>\Delta_0$. Since the left domain in Fig. \ref{model} is free of
the Zeeman field, the quasiparticle spectrum for plane waves is simply
given by $E=\sqrt{k^{2}+\Delta_{0}^{2}}$. For the bound state of
$E=0$, Hamiltonian (\ref{bdg}) can be solved  to yield
$(u_{\uparrow(\downarrow)},v_{\uparrow(\downarrow)})^{T}=e^{\pm
x\Delta_{0}}(1,\pm ie^{-i\varphi_{L}})^{T}$ for $x<0$. The
wavefunction proportional to $e^{-x\Delta_{0}}$ is invalid because
it diverges as $x\rightarrow-\infty$, so that we have
$(u_{\uparrow(\downarrow)},v_{\uparrow(\downarrow)})^{T}=e^{
x\Delta_{0}}(1,ie^{-i\varphi_{L}})^{T}$.
For the zero-energy mode of $x>0$, one finds that solutions are
decoupled into the spin-up and spin-down components in
Eq.(\ref{bdg}), yielding
$(u_{\uparrow},v_{\uparrow})^{T}=Ae^{x\lambda_{+}}(1,i)^{T}+B
e^{x\lambda_{-}}(1,-i)^{T}$ and
$(u_{\downarrow},v_{\downarrow})^{T}=Ce^{-x\lambda_{+}}(1,-i)^{T}+D
e^{-x\lambda_{-}}(1,i)^{T}$ with
$\lambda_{\pm}=(h\pm\Delta_{0})/\hbar v_F$. Since the solution in
spin-up component  diverges as $e^{x\lambda_{\pm}}$
($\lambda_{\pm}>0$ as $h>\Delta_0$) for $x\rightarrow\infty$, we
have $A=B=0$, and coefficients $C$ and $D$ are determined by the
boundary condition of wavefunction continuousness at $x=0$. As a
result, after gauging away the phase factor, the wavefunction of $E=0$
in the left junction is obtained as
\begin{equation}\label{bcl}
|\psi_{L}\rangle=e^{x\Delta_{0}}(0,e^{i(\frac{\varphi_{L}}{2}-\frac{\pi}{4})},
e^{-i(\frac{\varphi_{L}}{2}-\frac{\pi}{4})},0)^{T}
\end{equation}
for $x<0$ and
\begin{eqnarray}\label{s1}
|\psi_{L}\rangle&=&e^{-x\lambda_{+}}\sin\frac{\varphi_{L}}{2}(0,e^{\frac{\pi}{4}i},e^{-\frac{\pi}{4}i},0)^{T}\nonumber\\
&\quad&+e^{-x\lambda_{-}}\cos\frac{\varphi_{L}}{2}(0,e^{-\frac{\pi}{4}i},e^{\frac{\pi}{4}i},0)^{T}
\end{eqnarray}
for $x>0$. It can be readily shown that the quasiparticle operator
defined above  as $\gamma_{L}=\langle\psi_{L}|\Psi\rangle$ satisfies
relation $\gamma_{L}=\gamma_{L}^{\dagger}$, and so there is an MF in
spin-down component localized at $x=0$.

The same approach can be applied to the right junction at $x=L$. It
is found that there exists an MF in spin-up component located at $x=L$,
whose wavefunctions are obtained as
\begin{eqnarray}\label{s3}
|\psi_{R}\rangle&=&e^{(x-L)\lambda_{+}}\sin\frac{\varphi_R}{2}
(e^{\frac{\pi}{4}i},0,0,-e^{-\frac{\pi}{4}i})^T\nonumber\\
&\quad&+e^{(x-L)\lambda_{-}}\cos\frac{\varphi_R}{2}(e^{-\frac{\pi}{4}i},0,0,-e^{\frac{\pi}{4}i})^T
\end{eqnarray}
for $x<L$, and
\begin{equation}\label{bcr}
|\psi_{R}\rangle=e^{-(x-L)\Delta_{0}}(e^{(\frac{1}{2}\varphi_{R}-\frac{\pi}{4})i},0,0,
-e^{-(\frac{1}{2}\varphi_{R}-\frac{\pi}{4})i})^T
\end{equation}
for $x>L$. Obviously, $\gamma_{R}=\langle\psi_{R}|\Psi\rangle$
satisfies $\gamma_{R}=\gamma_{R}^{\dagger}$. Away from each
interface, the MF has two decay lengths: $1/\lambda_+$ and
$1/\lambda_-$, which are closely related to two energy gaps
$\Delta_{\pm}=\mid h\pm\Delta_{0}\mid$. Which one dominates the
decay rate of the MF depends upon the phase difference of the
junction. For example, if $\varphi_L=2N\pi$ with $N$ an arbitrary
integer, the left MF decays as $e^{-x\lambda_{-}}$ for $x>0$, while
at $\varphi_{L}=(2N+1)\pi$, it decays as $e^{-x\lambda_{+}}$ for
$x>0$.

\begin{figure}[hbtp!]
\includegraphics[scale=0.12]{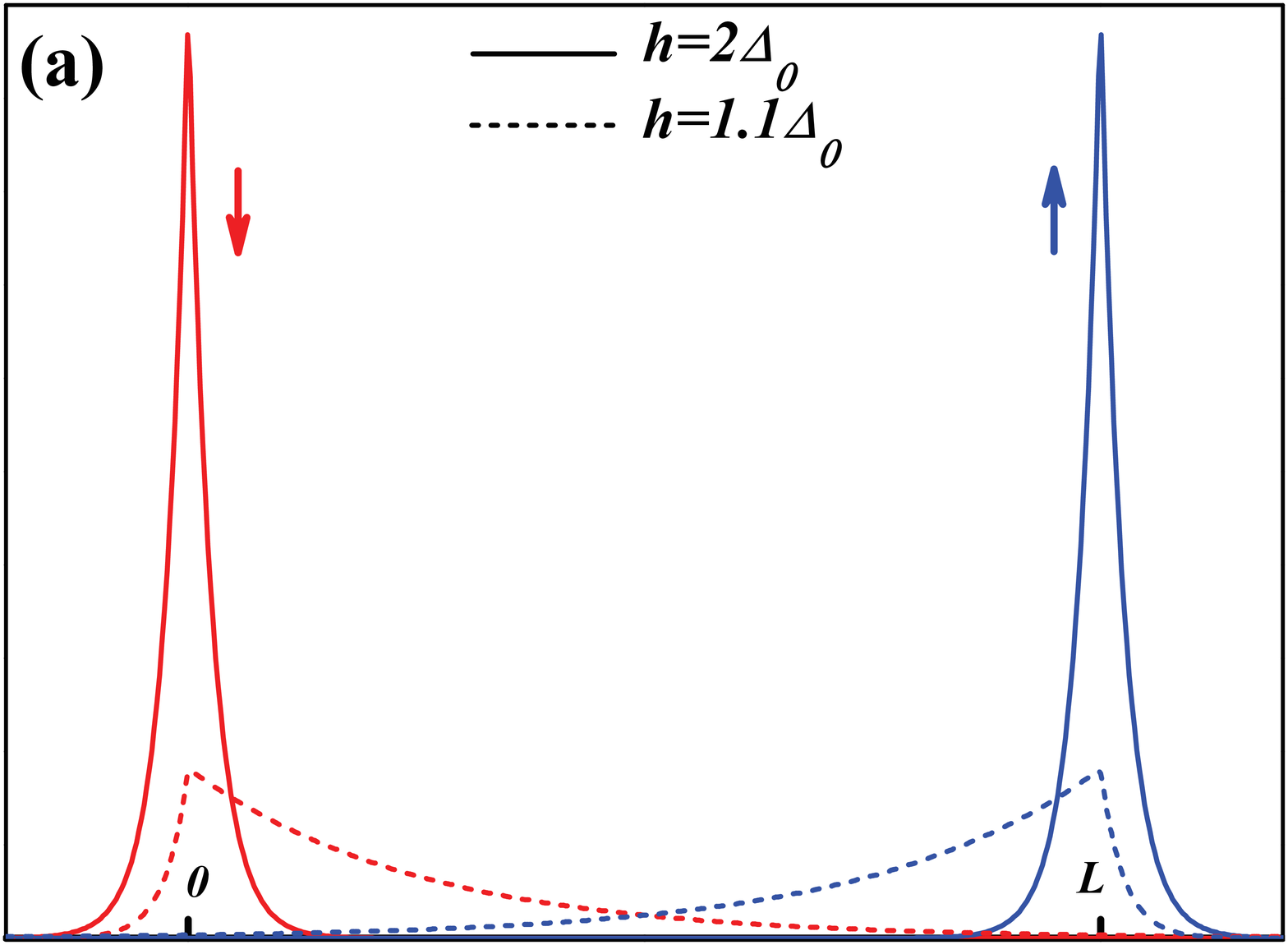}\includegraphics[scale=0.12]{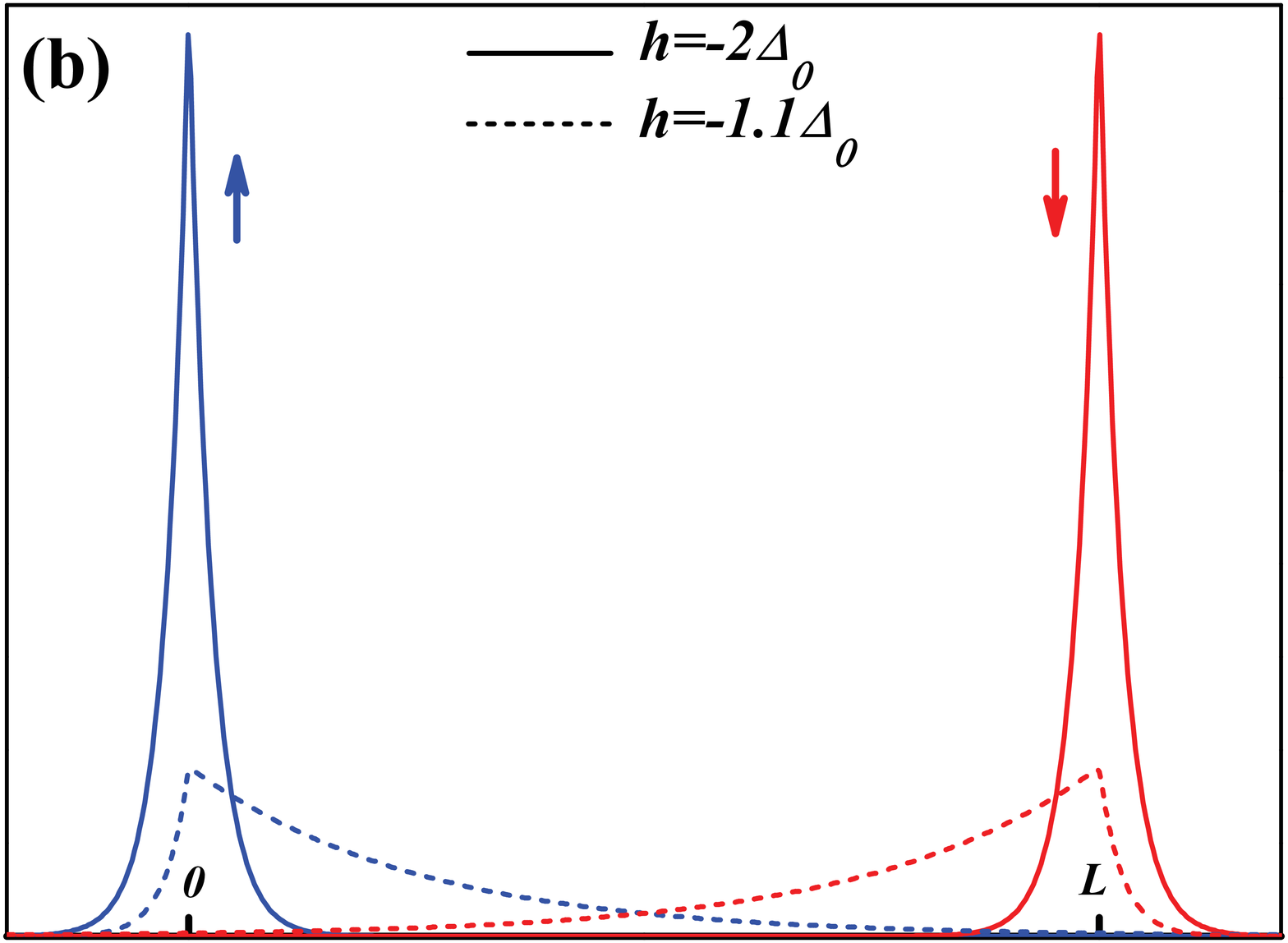}
\caption{(Color online) Space distribution of probability density $
\mid u\mid^2$ of the MFs for different parities (red and blue lines)
with $\varphi_{L}=\varphi_{R}=0$. The exchange field is
taken to be $h=2\Delta_{0}$ (solid line) and $h=1.1\Delta_{0}$
(dashed line) in (a), and $h=-2\Delta_{0}$ (solid line) and
$h=-1.1\Delta_{0}$ (dashed line) in (b).} \label{F2}
\end{figure}

In Fig. \ref{F2},
we plot zero-energy quasiparticle probability $\mid u\mid^2$ for $h
>\Delta_{0}$ (a) and  $h < -\Delta_{0}$ (b) as
$\varphi_{L}=\varphi_{R}=0$. As shown in Fig.\ 2(a), there are  two
MFs localized at $x=0$ and $x=L$ for $h>\Delta_0$. As the magnitude
of the Zeeman field is turned down, the MFs become more and more
extended. And when $|h|<\Delta_0$, the two MFs are annihilated on
the junctions and fused. Such a fusion of MFs arises
from the sign reverse of $\lambda_-$ at $h=\Delta_0$ ($\lambda_+$ at
$h=-\Delta_0$) due to the closing of energy gap $\Delta_-$
($\Delta_+$). For $\mid h\mid<\Delta_{0}$, since $\lambda_+>0$ and
$\lambda_-<0$, the second terms in Eqs.~(\ref{s1}) and (\ref{s3}) would
diverge as $x\rightarrow\infty$, and so the solution for the MFs
would be an unphysical result. Therefore, there is no MF for $\mid
h\mid<\Delta_{0}$, and there appears a topological transition  at
$\mid h\mid=\Delta_{0}$ from the MF phase to the trivial one without
MF. If the direction of the Zeeman field ($h>\Delta_0$) is reversed,
we have $h<-\Delta_{0}$ and so $\lambda_{\pm}<0$, with the result being
shown in Fig.\ 2(b). In this case, using the same procedure of
calculation, the wavefunctions on both sides of each interface at
$x=0$ or $x=L$ can be obtained as follow. The MF located at $x=0$ is now
obtained in the spin-up component, and its wavefunction is given by
$|\psi_{L}^{'}\rangle=e^{x\lambda_{-}}\sin\frac{\varphi_{L}}{2}
(-e^{-\frac{\pi}{4}i},0,0,e^{\frac{\pi}{4}i})^T
+e^{x\lambda_{+}}\cos\frac{\varphi_{L}}{2}(e^{\frac{\pi}{4}i},0,0,-e^{-\frac{\pi}{4}i})^T$
for $x>0$. Compared with Eq. (\ref{s1}), one finds that the
direction reverse of the exchange field leads to that the MF in the spin-down
component is replaced by an MF in the spin-up component, accompanied with
an exchange of $\exp(-x\lambda_{\pm})$ and $\exp(x\lambda_{\mp})$
due to the sign reverse of $\lambda_{\pm}$. Similarly, the wavefunction
 of the MF located at $x=L$ is obtained in the spin-down component as
$|\psi_{R}^{'}\rangle=-e^{-(x-L)\lambda_{-}}\sin\frac{\varphi_{R}}{2}
(0,e^{-\frac{\pi}{4}i},e^{\frac{\pi}{4}i},0)^T+e^{-(x-L)\lambda_{+}}\cos\frac{\varphi_{R}}{2}
(0,e^{\frac{\pi}{4}i},e^{-\frac{\pi}{4}i},0)^T$
for $x<L$. It can also be obtained by performing a symmetry
operation to recover the solution of a reversed exchange field.
Hamiltonian (\ref{h}) satisfies $UH(h)U^{\dagger}=H(-h)$ with
$U=\hat{\mathcal{R}}\hat{\mathcal{O}}$ where $\hat{\mathcal{R}}$ is
a spin-rotation of $\pi$ around $\hat{x}$ and $\hat{\mathcal{O}}$ is
the central inversion with $x\rightarrow L-x$. It is obvious that
$|\psi_{L,R}^{'}\rangle\sim U|\psi_{L,R}\rangle$.

The above discussion  is suitable to the case of
$\varphi_{L,R}\neq(2N+1)\pi$.
\begin{figure}[hbtp!]
\includegraphics[scale=0.12]{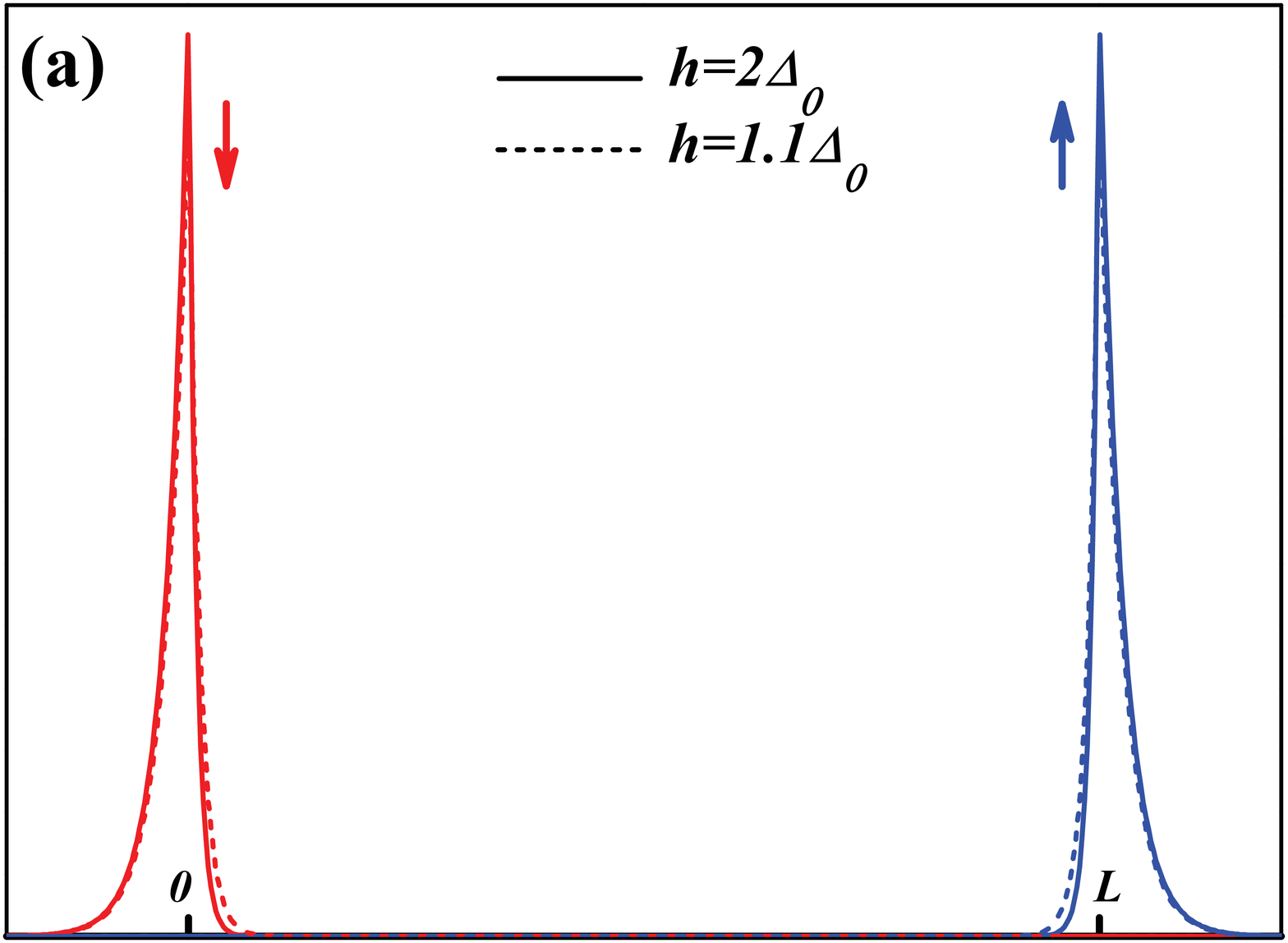}\includegraphics[scale=0.12]{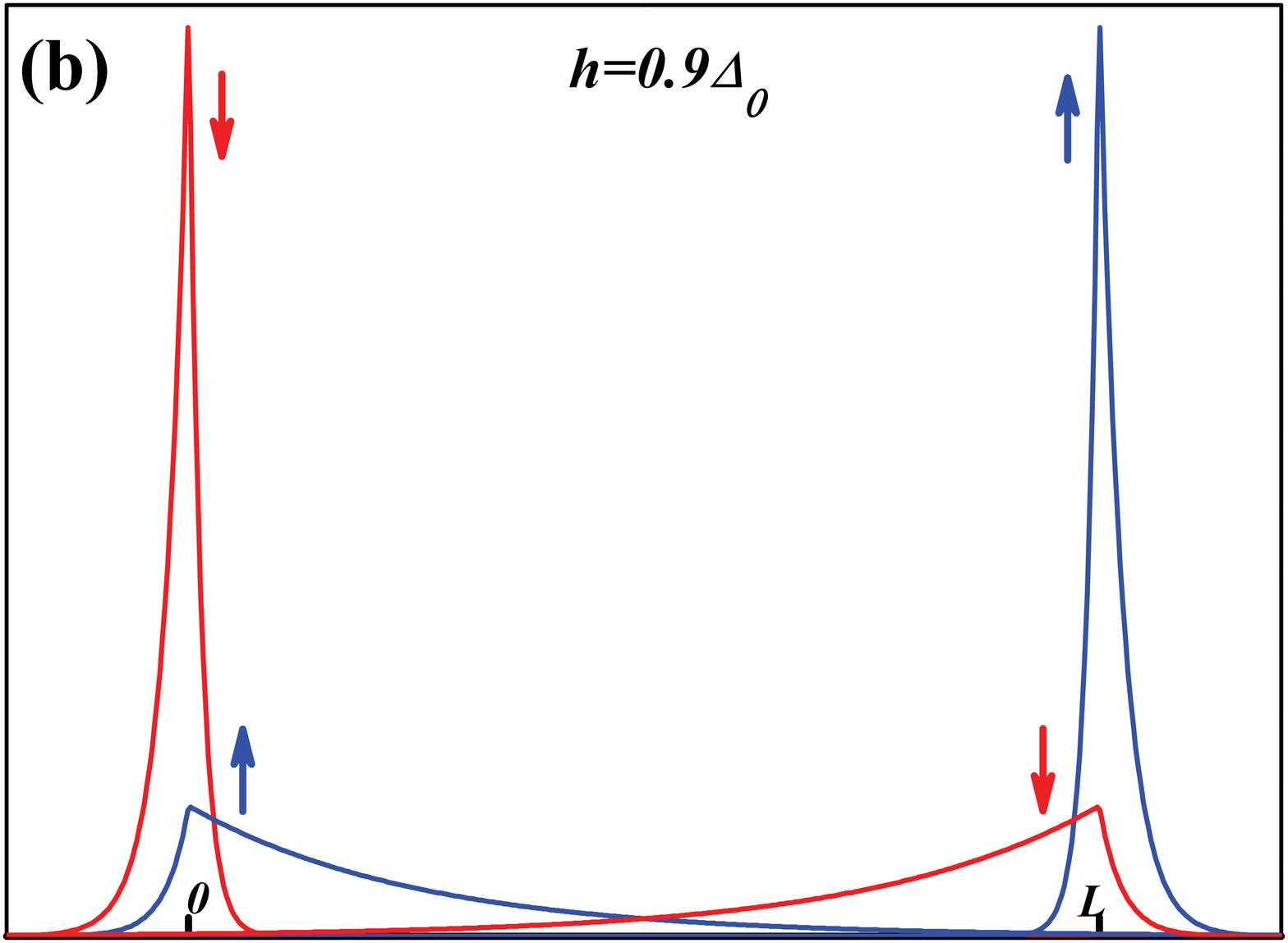}\\
\includegraphics[scale=0.12]{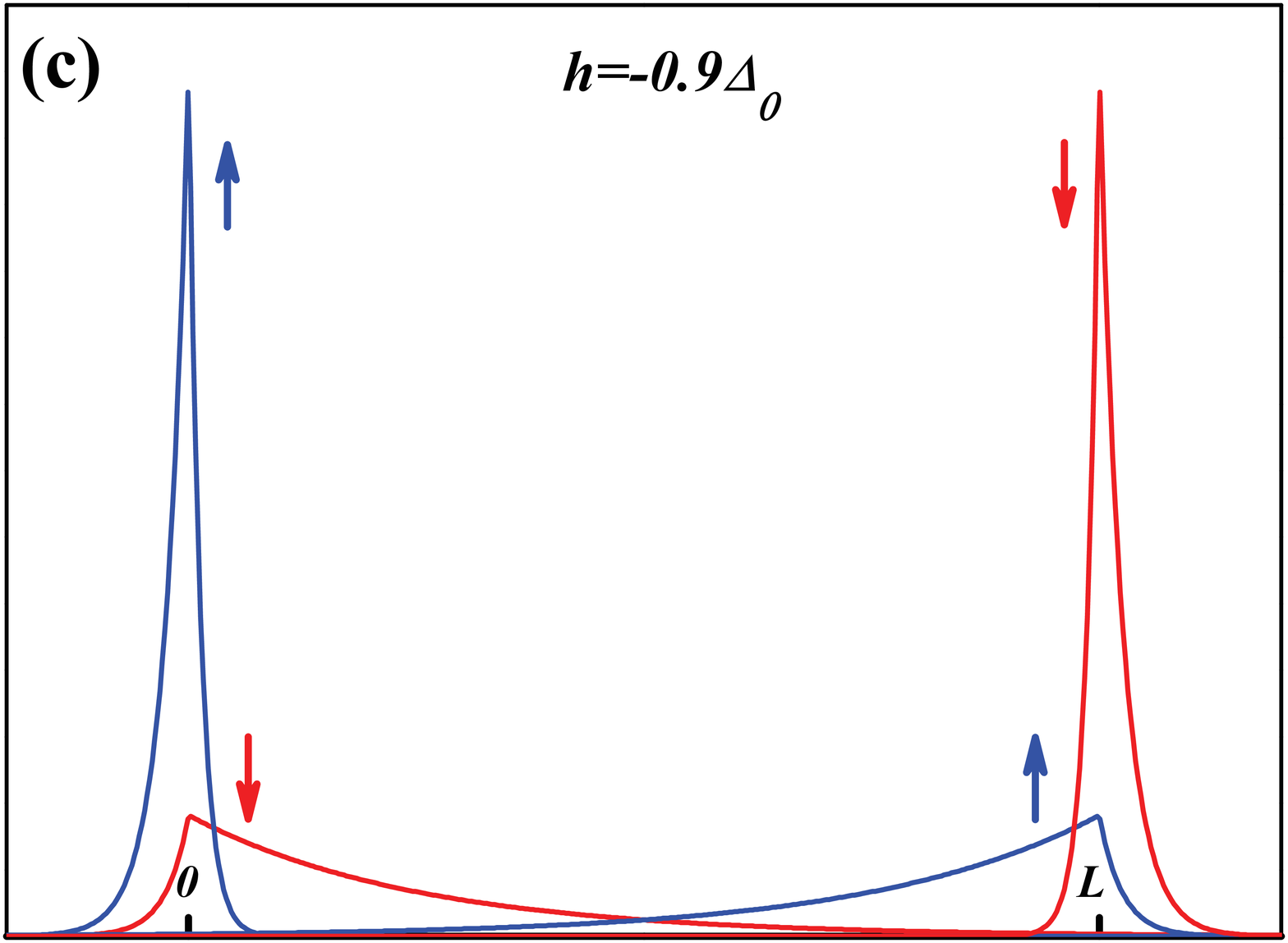}\includegraphics[scale=0.12]{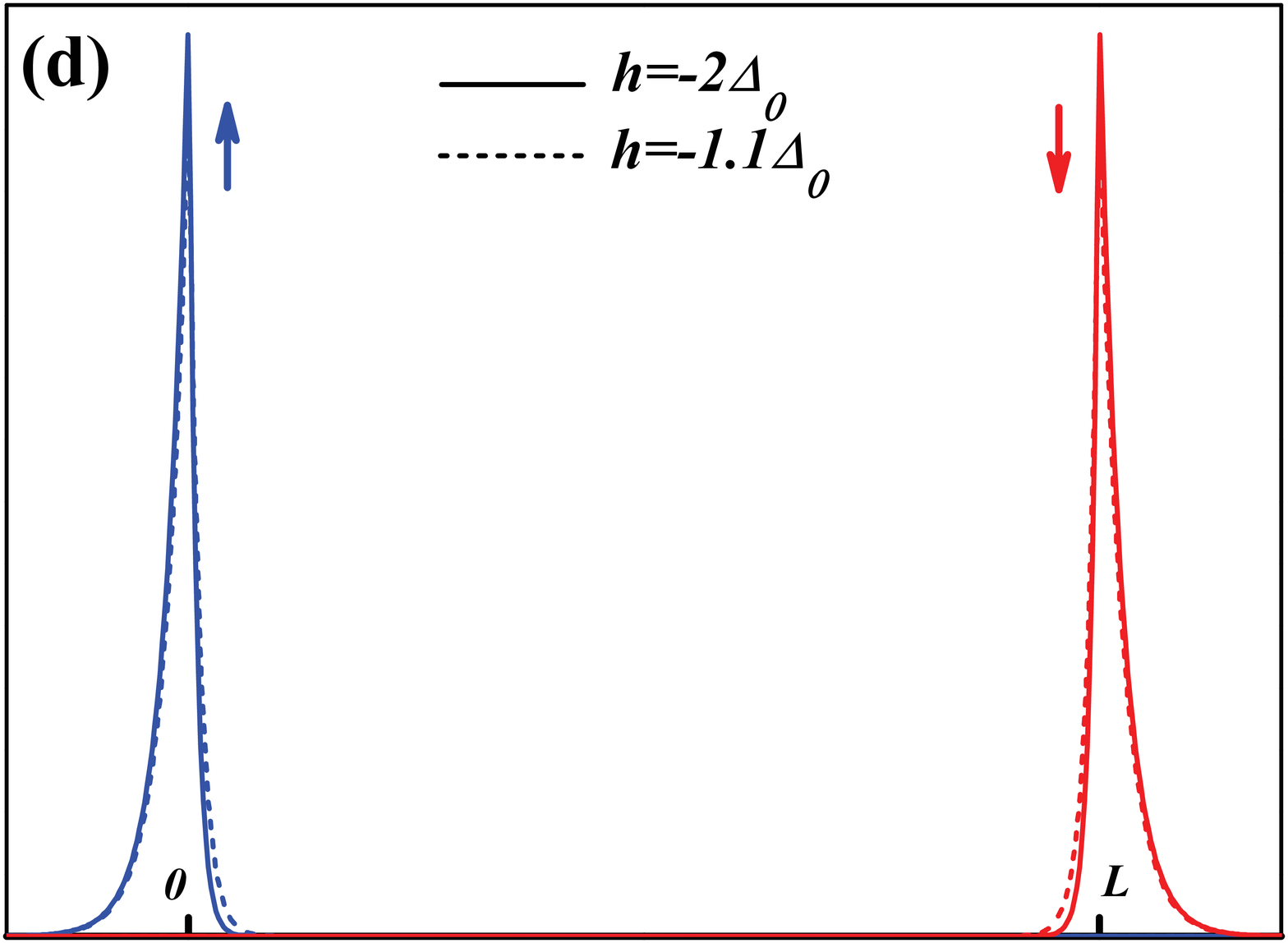}
\caption{(Color online) Space distribution of $ \mid u\mid^2$ for
different parities (red and blue lines) with
$\varphi_{L}=\varphi_{R}=\pi$. The exchange field is
taken to be $h=2\Delta_{0}$ (solid line) and $h=1.1\Delta_{0}$
(dashed line) in (a),  $h=0.9\Delta_{0}$  in (b), $h=-0.9\Delta_{0}$
in (c), and $h=-2\Delta_{0}$ (solid line) and $h=-1.1\Delta_{0}$
(dashed line) in (d). } \label{F3}
\end{figure}
For $\varphi_{L,R}=(2N+1)\pi$,  the second terms of Eqs. (\ref{s1})
and (\ref{s3}) vanish, and the calculated results for zero-energy
quasiparticle distributions are plotted in Fig. \ref{F3}.  It is
found that there exist still two MFs on the junctions for
$h>\Delta_{0}$. The essential difference is that the MF peaks have
merely a very slight extension with decreasing $ h$, as shown in Fig.
\ref{F3}(a) and (b). At the same time,  another pair of MFs are
generated just after closing energy gap $\Delta_{-}=|h-\Delta_{0}|$
so as to form two Dirac fermions at $x=0$ and $x=L$. Such a novel
behavior can be understood by the following argument. Taking the
left junction for example again, the MF state in Eq.\ (7) decays as
 $e^{-x\lambda_{+}}$ for $x>0$. For $|h|<\Delta_0$, regardless
of the closing of gap $\Delta_{-}$, $\Delta_+=h+\Delta_0$ makes the
original MF survive, for this MF is protected only by gap
$\Delta_+$. More interestingly, there appears an additional MF in
the spin-up component at $x=0$, whose wavefunction is proportional to
$e^{x\lambda_{-}}$ with $\lambda_- <0$. This MF must be combined
with the original one, forming a Dirac fermion. Since a change of
the phase for a superconductor will not close energy gap of the
bulk, the present phase with the Dirac fermion located at the junction
for $\varphi_{L,R}=(2N+1)\pi$ is topologically equivalent to that
without Dirac fermion there for $\varphi_{L,R}\neq(2N+1)\pi$. We
wish to point out here that the MF in the spin-up component at $x=0$
exists only for $\Delta_0>h$, and annihilates for $h>\Delta_0$ due
to the sign reverse of $\lambda_-$ at $h=\Delta_0$. The same argument is
also applied to the right junction. As a result, a pair of Dirac
fermions are formed in Fig. \ref{F3} (b). For $h<0$, the particle
distribution is shown in Figs. \ref{F3}(c) and (d). With increasing
the magnitude of $h<0$, the topological phase transition occurs once
again at $h=-\Delta_0$.

\begin{figure}[hbtp!]
\includegraphics[scale=0.12]{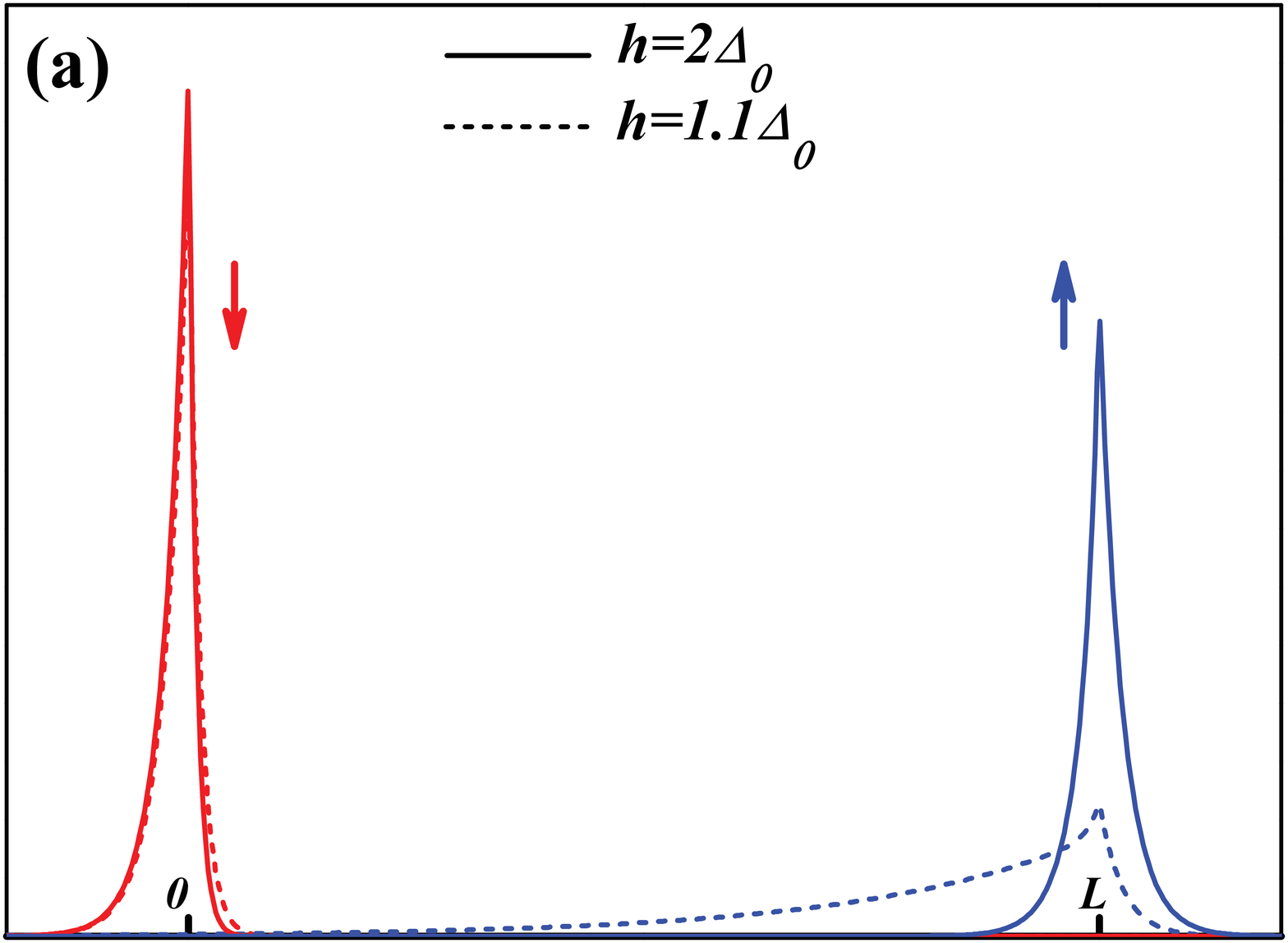}\includegraphics[scale=0.12]{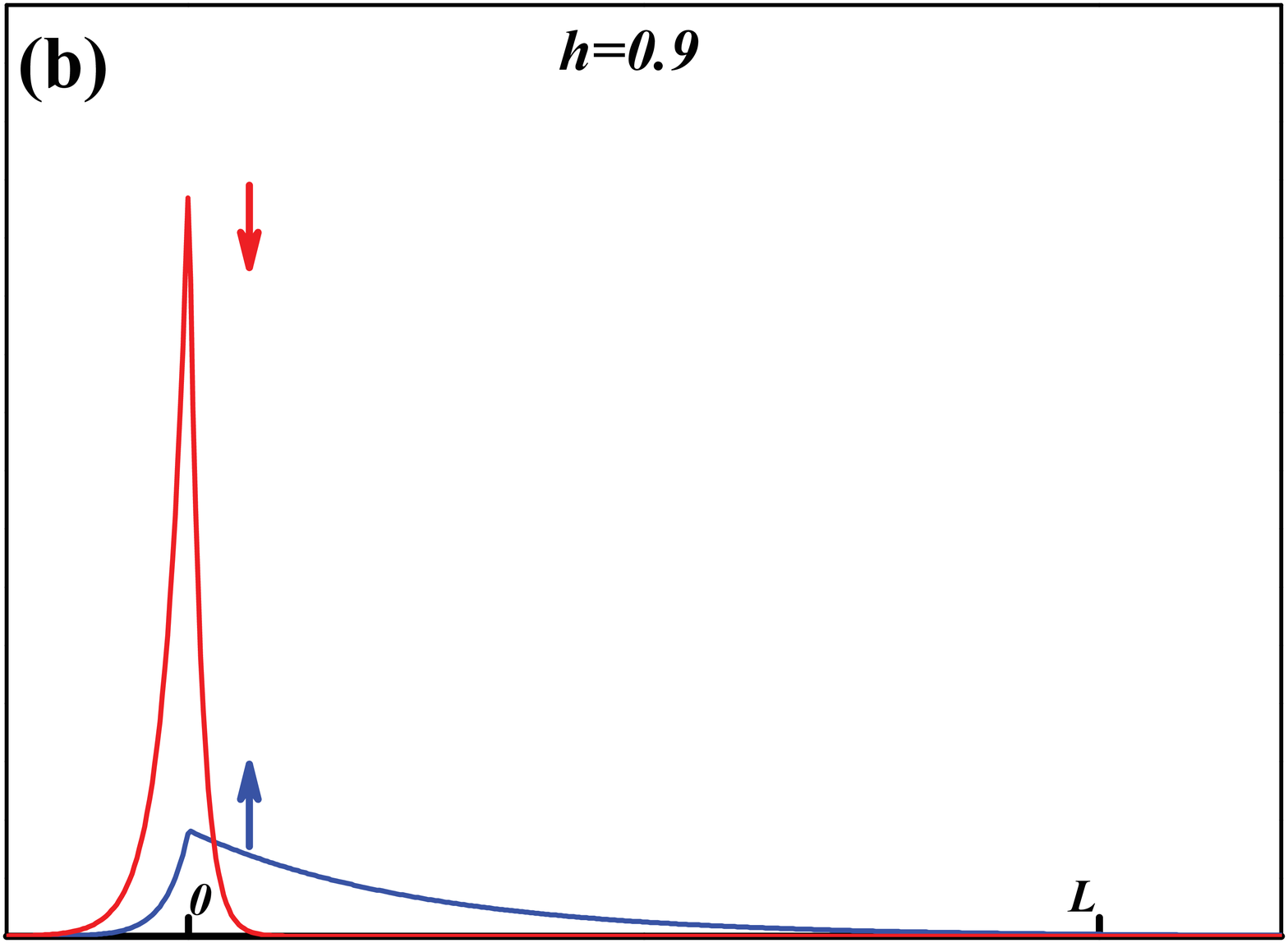}\\
\includegraphics[scale=0.12]{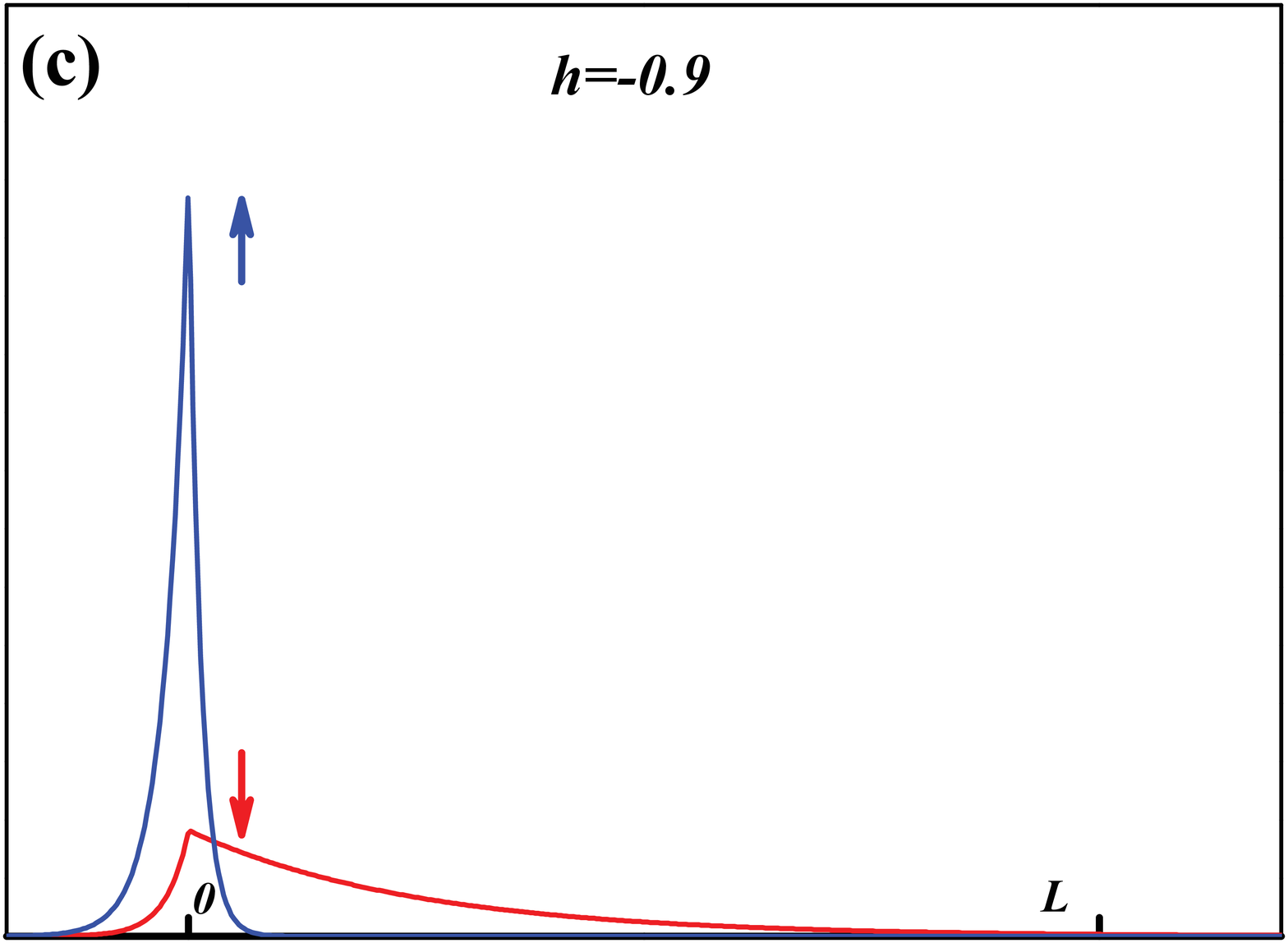}\includegraphics[scale=0.12]{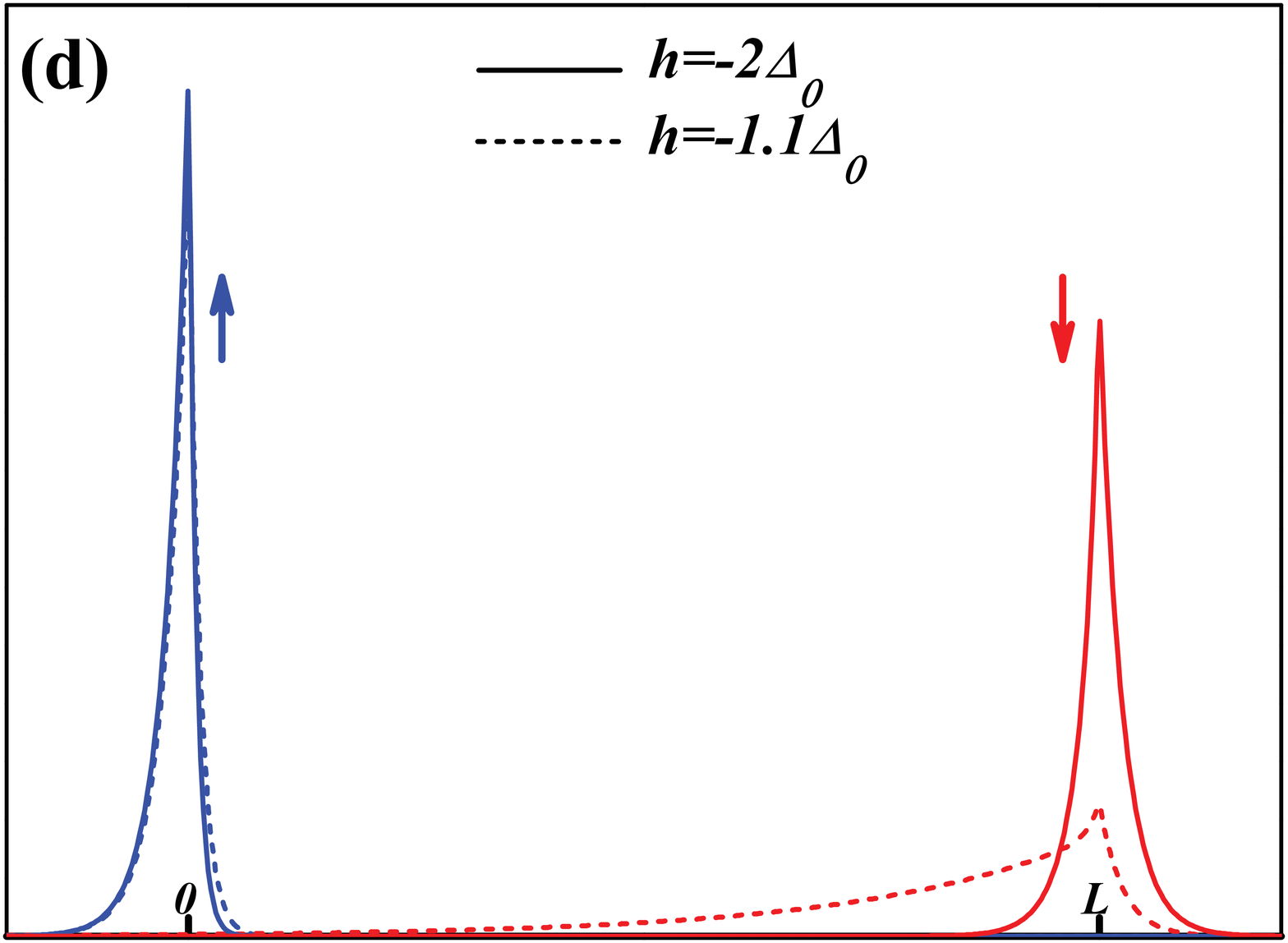}
\caption{(Color online) Space distribution of $ \mid u\mid^2$ for
different parities (red and blue lines) with $\varphi_{L}=\pi$ and $\varphi_{R}=\pi/3$. The exchange field is set
$h=2\Delta_{0}$ (solid line) and $h=1.1\Delta_{0}$ (dashed line) in
(a),  $h=0.9\Delta_{0}$  in (b), $h=-0.9\Delta_{0}$  in (c), and
$h=-2\Delta_{0}$ (solid line) and $h=-1.1\Delta_{0}$ (dashed line)
in (d). } \label{F4}
\end{figure}

For $\varphi_{L}=(2N+1)\pi$ and $\varphi_{R}\neq(2N+1)\pi$, the
situation is also interesting, and the space distribution of
$|u|^{2}$ with different Zeeman fields is plotted in Fig. \ref{F4}.
For $h>\Delta_{0}$ in Fig. \ref{F4}(a), the right MF extends rapidly
into the bulk with decreasing $h$, whereas the left MF remains
almost unchanged. As $h$ is less than $\Delta_0$, the MF on the
right junction is annihilated and at the same time another MF is
created on the left junction, as shown in Fig. \ref{F4}(b). This
evolution is equivalent to the process that the MF on the right
junction is driven to the left junction and two MFs there are
combined to form one Dirac fermion. As $h$ is reversed, the two MFs
that combine into one Dirac fermion located at $x=0$ exchanges their
magnitudes [see Fig. \ref{F4}(c)]; and as $h\leq -\Delta_0$, an MF
moves back to the right junction and the system reenters the
topological phase [see Fig. \ref{F4}(d)]. The underlying physics has
been discussed above, and will not be repeated here. The evolution
from MFs to Dirac fermions in Fig. \ref{F4} can be used to detect
the physical state of MFs. We can drive them to form Dirac fermions
for detection and initialize MFs between subspaces of different
parities.

The MFs on the right and left junctions are coupled with each other,
i.e., $\langle\psi_L|\mathcal{H}_{BdG}|\psi_{R}\rangle\neq0$. For
$h>\Delta_0$, $|\psi_{L}\rangle$ and $|\psi_{R}\rangle$ have been
given by Eqs. (\ref{bcl}) and (\ref{s1}), and (\ref{s3}) and
(\ref{bcr}), respectively. It can be shown that such a coupling
depends to a great degree upon phase differences of the two
junctions, and it will vanish if the following condition is
satisfied,
\begin{equation}\label{tan}
\cot\frac{\varphi_{L}-\varphi}{2}\cot\frac{\varphi-\varphi_{R}}{2}=e^{-2L/\xi},
\end{equation}
where $\xi=\hbar v_F/\Delta_0$ is the superconducting coherent
length. In this case, we have a pair of zero-energy MFs decoupled
exactly, such as those in Figs.\ (3) and (4) where $\varphi_{L}=\pi$
and/or $\varphi_{R}=\pi$ and $L\gg\xi$. If condition (8) is not
satisfied, the coupling will make the MFs have a small departure
from zero energy, proportional to $e^{-L/\xi}$.

In summary, we have shown that the edge state of a 2D topological
insulator in the proximity with $s$-wave superconductors and under a
vertical Zeeman field may accommodate MFs. The MFs can be
manipulated by tuning the phase differences of the junction to either be
fused or form one Dirac fermion on the $\pi$-junction.
As the exchange field becomes less than the superconducting
order parameter, one MF can be driven from a non-$\pi$-junction to a
$\pi$-junction, forming one Dirac fermion on the $\pi$-junction.
There exhibits a topological phase transition at $\mid h\mid =
\Delta_{0}$. For $\mid h\mid < \Delta_{0}$, there are three topologically
equivalent phases with different phase differences: one and two
Dirac fermions localized on the $\pi$-junctions, and none.
We have also illustrated a condition that
decouples completely the two MFs at two junctions regardless of the
magnitude of the exchange field.

\textbf{Acknowledgment} We would like to thank Shi-Liang Zhu for
helpful discussions. This work is supported by the GRF (HKU7058/11P) and CRF (HKU-8/11G) of the RGC of Hong Kong,
the State Key Program
for Basic Researches of China under Grant Nos.  2011CB922103 and
2010CB923400 (DYX), by the National Natural Science Foundation of
China under Grant Nos. 10874066, 11074110 (LS), 11174125, 11074109,
and 91021003 (DYX), and by a project funded by the PAPD of Jiangsu
Higher Education Institutions.

\end{document}